\documentclass[12pt]{article}

\usepackage{epsfig,latexsym,amsfonts,amsmath,amsthm,amssymb,amsbsy,multirow,slashed,wasysym,textcomp,subfigure,hyperref,wrapfig,datetime,comment,mathtools,cancel,cite,mathrsfs,footmisc,mathrsfs}

\usepackage{scalerel}
\hypersetup{colorlinks=true, linkcolor=green, citecolor=cyan, linktoc=page}
\makeatother

\usepackage{stackengine,wasysym}
\usepackage{booktabs}

\usepackage[font=small,skip=-15pt]{caption} 
\def\bee#1\eee{\begin{align}#1\end{align}}

\def\ee{\end{equation}}
\def\bea{\begin{eqnarray}}
\def\eea{\end{eqnarray}}
\newcommand{\beq}{\begin{eqnarray}}
\newcommand{\eqq}{\end{eqnarray}}
 \newcommand{\badat}{\begin{alignedat}}
 \newcommand{\eadat}{\end{alignedat}}

\newcommand{\eal}[1]{\be \begin{aligned} #1 \end{aligned}\end{equation}} 
\newcommand{\eqn}[1]{\be #1 \end{equation}} 
\newcommand{\eqa}[1]{\bea  #1\end{eqnarray}}

\long\def\new#1\endnew{{\bf #1}}		
\long\def\del#1\enddel{}

\def\del{\partial}

\usepackage{array}
\usepackage{tikz}
\usepackage{multirow,multicol}
\usepackage{xcolor}
\usepackage{dirtytalk}
\usepackage{mathrsfs} 
\definecolor{oldmauve}{rgb}{0.4, 0.19, 0.28}
\definecolor{pansypurple}{rgb}{0.47, 0.09, 0.29}
\definecolor{burgundy}{rgb}{0.5, 0.0, 0.13}
\definecolor{carminepink}{rgb}{0.92, 0.3, 0.26}
\definecolor{blue(pigment)}{rgb}{0.2, 0.2, 0.6}
\definecolor{darkseagreen}{rgb}{0.56, 0.74, 0.56}
\definecolor{darkspringgreen}{rgb}{0.09, 0.45, 0.27}
\definecolor{ceruleanblue}{rgb}{0.16, 0.32, 0.75}

 \usepackage[framemethod=tikz]{mdframed}
\definecolor{mycolor}{rgb}{0.122, 0.435, 0.698}
            
\newmdenv[innerlinewidth=0.5pt, roundcorner=4pt,linecolor=mycolor,innerleftmargin=6pt,
innerrightmargin=6pt,innertopmargin=6pt,innerbottommargin=6pt]{bluebox}

\newcommand{\be}{\begin{eqnarray}}
\newcommand{\en}{\end{eqnarray}}

\numberwithin{equation}{section} 

\begin{document}

\begin{titlepage}
  \thispagestyle{empty}

  \begin{center}  
  

{\LARGE\textbf{Stability of Extremal Black Holes and  Weak Cosmic Censorship Conjecture in Kiselev Spacetime}}\\
\vspace{0.2cm}

\vskip1cm
Ankit Anand$^\star$\footnote{\fontsize{8pt}{10pt}\selectfont\ \href{mailto:ankitanandp94@gmail.com}{ankitanandp94@gmail.com}}, Anshul Mishra $^{\dagger}$\footnote{\fontsize{8pt}{10pt}\selectfont\ \href{mailto:anshulmishra2025@gmail.com}{anshulmishra2025@gmail.com}} and Phongpichit Channuie$^{*,\ddag}$\footnote{\fontsize{8pt}{10pt}\selectfont\ \href{mailto:: phongpichit.ch@mail.wu.ac.th}{channuie@gmail.com}}

\vskip0.5cm

\normalsize
\medskip

$^\star$\textit{Physics Division, School of Basic and Applied Sciences, Galgotias University, Greater Noida 203201, India.}
 
$^\dagger$\textit{Physics Department, Shyamlal Saraswati Mahavidyalaya, Shikarpur,Bulandshahr 203395, India.}

$^*$\textit{School of Science, Walailak University, Nakhon Si Thammarat, 80160, Thailand.\\
$^\ddag$ College of Graduate Studies, Walailak University, Nakhon Si Thammarat, 80160, Thailand.}


\vskip1cm

\vspace{1.5cm}
\begin{abstract} 

In this study, we investigate the Weak Gravity Conjecture (WGC) and Weak Cosmic Censorship Conjecture (WCCC) for a quantum-corrected Reissner-Nordström Anti-de Sitter (RN-AdS) black hole embedded in Kiselev spacetime. By making small perturbations to the action and using WGC, we investigate the stability of black holes and predict the existence of lighter particles in the spectrum. Using the scattering of a charged scalar field, we study the WCCC. We verify under certain conditions on the temperature of the black hole, the second law holds for near-extremal black holes. Finally, we demonstrate that the WCCC holds for both extremal and near-extremal black holes.

\end{abstract}

\vskip2cm

\end{center}

\end{titlepage}

\section{Introduction}\label{Introduction}

Black hole thermodynamics represents a refined domain within physics that investigates the intricate relationship between thermodynamic laws and the characteristics of black holes. These enigmatic entities, intrinsically linked to classical parameters like horizon area and surface gravity, possess thermodynamic properties such as entropy and temperature\cite{Hawking:1975vcx, Bardeen:1973gs, PhysRevD.7.2333}. The study of black hole thermodynamics necessitates a synthesis of general relativity, quantum mechanics, and thermodynamic principles, culminating in a holistic framework for understanding black holes. The impetus for delving into black hole thermodynamics stems from various compelling factors: A notable parallel exists between black hole mechanics principles and those of classical thermodynamics. The groundbreaking discovery that black holes emit thermal radiation, thereby possessing a finite temperature\cite{Hawking:1976de}, has profound implications for our understanding of these cosmic phenomena. This principle posits that the informational content of a spatial region correlates with its boundary area rather than its volume, further enriching the discourse on black hole properties.  This theoretical framework establishes a connection between gravitational theories in anti-de Sitter (AdS) space and conformal field theories (CFT) \cite{Maldacena:1997re} on their boundaries, providing deep insights into the nature of black holes.

\quad The Swampland Program \cite{Palti Eran:2019, vanBeest:2021lhn} represents a transformative initiative in theoretical physics aimed at distinguishing effective low-energy theories incompatible with quantum gravity from those that are viable. This approach starkly contrasts with the string theory landscape, which encompasses theories that align with quantum gravity, a coherent framework for quantum mechanics and gravity. The Swampland, by contrast, includes theories that may initially appear consistent but ultimately lack a viable ultraviolet (UV) completion when gravity is taken into account. A significant tool in this program is the Weak Gravity Conjecture (WGC), which suggests that in any consistent theory of quantum gravity, gauge forces must always exceed gravitational forces for certain particles, affirming gravity as the weakest force \cite{Arkani-Hamed Nima, Harlow}. This conjecture not only helps to separate out incompatible theories but also serves as an upper mass limit, reinforcing the program’s principles. Initiated by Cumrun Vafa \cite{Vafa:2005ui}, the Swampland Program seeks to establish boundaries within the quantum gravity landscape by identifying universal criteria shared by all theories capable of achieving a gravitational UV completion \cite{Ooguri:2006in, Arkani-Hamed:2006emk, Heidenreich:2016aqi, Palti:2017elp, Odintsov:2020zkl}. This conjecture has profound implications across various domains of physics and mathematics, suggesting that gravity should invariably be the weakest force in any consistent quantum gravity theory. The WGC has been extensively reviewed and is supported by compelling evidence from examples within string theory. The Swampland Program and the WGC are pivotal in our ongoing quest to comprehend the fundamental principles governing our universe at the quantum level. Their implications resonate through particle physics, cosmology, general relativity, and mathematics, guiding researchers in exploring the limits and possibilities of theoretical physics. To examine specific applications of the Swampland Program in contexts such as inflation, black holes, thermodynamics, and black branes \cite{Liu:2021diz, Liu:2022myp, g, Alipour:2023css, Alipour:2023yiz, Sadeghi:2023cui, Schoneberg:2023lun, Kadota:2019dol, Oikonomou:2020oex, NooriGashti:2022xmf, Das:2019hto, Yuennan:2022zml, Bedroya:2019snp, Sadeghi:2023cpd, v, w, x, Guleryuz:2021zik, Osses:2021snt, aa, Brahma:2019unn, Brandenberger:2021pzy, Sadeghi:2022tzd, Geng:2019bnn, Gashti:2022hey, gg, hh, Agrawal:2018own, kk, ll, mm, Odintsov:2017hbk, Shokri:2021zqw, Sadeghi:2021plz, Shokri:2021hjs, Sadeghi:2020xtc, Yuennan:2022vdd, uu, Kinney:2018kew, Kinney:2021hje, Yu:2018eqq, NooriGashti:2021nox,  Sadeghi:2022wgx, Sadeghi:2022xcr, Kolb:2021nob, Sadeghi:2021zjb, Cho:2023koe, Nakayama:2015hga, Aalsma:2020duv, Sadeghi:2023cxh, Heidenreich:2021yda, Loges:2019jzs}.

\quad  The WGC and the WCCC represent two distinct yet profound concepts in theoretical physics, each addressing unique facets of fundamental physics. While both conjectures incorporate the term \say{weak} and are concerned with essential aspects of gravity and spacetime, they diverge significantly in their focus and implications. The WGC primarily examines the relative strengths of forces within a quantum gravity framework. In contrast, the WCCC pertains to the behavior of singularities and event horizons during gravitational collapse scenarios, as classical general relativity describes. This distinction invites speculation regarding potential connections between the WGC and cosmic censorship, particularly whether the WGC might influence scenarios involving cosmic censorship. About black holes, the WCCC posits that singularities should remain concealed behind event horizons, thereby preserving the causal structure of spacetime. The interplay between the WGC and WCCC is both subtle and complex. The Weak Cosmic Censorship Conjecture (WCCC) posits that singularities arising from gravitational collapse must invariably remain hidden behind the event horizons of black holes, preventing them from being observed by distant observers. While a comprehensive proof of WCCC remains elusive, it has nonetheless become a foundational principle in black hole physics. To scrutinize the validity of Penrose's conjecture, Wald introduced a thought experiment involving the injection of a particle with substantial charge or angular momentum into an extremal Kerr–Newman black hole. Another approach to testing the WCCC involves scattering a classical test field, initially suggested by Semiz and further refined by subsequent researchers \cite{JiZh20, GMZZ18, NiCL19, DuSe13, SeDu15, Duzt15}.

\quad Quantum corrections are crucial in resolving the singularity dilemma that arises within classical general relativity. The conventional Schwarzschild solution reveals a singularity at \( r = 0 \), where the curvature of spacetime becomes infinitely large. Researchers D.I. Kazakov and S.N. Solodukhin have delved into spherically symmetric quantum fluctuations, examining metric and matter fields. Their findings culminate in an effective two-dimensional dilaton gravity model, which modifies the Schwarzschild solution by transforming the classical singularity at \( r = 0 \) into a quantum-corrected region. This new zone possesses a minimum radius, \( r_{\text{min}} \), approximately equivalent to the Planck length, \( r_{\text{PL}} \), ensuring that scalar curvature remains finite. The significance of this alteration lies in its implication of a singularity-free, regular spacetime structure. This spacetime consists of two asymptotically flat regions connected by a hypersurface of constant radius, suggesting that quantum effects may smooth out the singularities predicted by classical general relativity.
Consequently, exploring quantum corrections has garnered substantial interest among researchers, inspiring investigations across various domains. These include The criticality and efficiency of black holes. The thermodynamics of a quantum-corrected Schwarzschild black hole in quintessence environments. Accretion processes onto a Schwarzschild black hole influenced by quintessence. Studies of quasinormal modes, scattering phenomena, shadows, and the Joule-Thomson effect. These inquiries not only enhance our understanding of black hole physics but also deepen our insights into the quantum-level structure of spacetime.

\quad  The paper is organized as follows. In Section \ref{sec:Kislev Spacetime}, we discuss Kiselev Spacetime and its thermodynamics. In Section \ref{Sec:Weak Gravity Conjucture}, we perturbed our metric, calculated the perturbed metric and perturbed thermodynamic variables, and evaluated the effect of the perturbation on the extremal black hole. On one side, we confirm WGC in our system; on the other, we use WGC to predict the stability of extremal black holes and the existence of lighter particles in the spectrum. In Section \ref{Sec:Charged Scalar Field}, we very briefly discuss the Charged Scalar Field and computed the change over infinitesimal time.In Section \ref{Sec:The Weak Cosmic Censorship Conjectures}, we tested the WCCC validation. Finally, in Section \ref{Sec:Conclusion}, we concluded what we have studied.

\section{Kiselev Spacetime}\label{sec:Kislev Spacetime}

\quad This section will comprehensively analyze the quantum-corrected AdS-RN black hole within Kiselev spacetime's thermodynamics. M. Visser has suggested that the Kiselev black hole can be extended to a spacetime comprising \( N \) components, where each component's relationship between energy and pressure is linear \cite{Kiselev:2002dx, Visser:2019brz}. Specifically, the spacetime for this generalized Kiselev black hole can be expressed with form proportional to $r^{-(3\omega + 1)}$\footnote{The value of $\omega$ decides the different scenarios, e.g., By $\omega = 1/3$ and \( c < 0 \), behaves like adding an electric charge to the theory, (\( \omega = -1 \)), results in the anti-de Sitter, \( \omega = -2/3 \), \( -1/6 \), results in the inclusion of a cosmological fluid term designed to model quintessence and (\( \omega = -4/3 \)), for the phantom dark energy .}. The parameter $a$ signifies a modification of the black hole mass due to quantum corrections \cite{Kazakov:1993ha}. It is an independent parameter, and when $a=0$, the metric simplifies to the AdS-Reissner-Nordström metric. In principle, $a$ can take any value as long as it remains less than the event horizon radius, as this reflects a small correction to the black hole metric. We will delve into the spacetime metric of the quantum-corrected charged AdS black hole enveloped by Kiselev spacetime. This metric is characterized by its spherical symmetry and is expressed as follows \cite{MoraisGraca:2021ife}
\begin{equation}\label{eq1}
ds^2= g_{\mu \nu} dx^\mu dx^\nu = - f(r)dt^2+f(r)^{-1}dr^2+r^2d\Omega^2,
\end{equation}
According to \cite{MoraisGraca:2021ife}, the function $f(r)$ is defined by the equation:
\begin{equation}\label{eq2}
\begin{split}
f(r)=-\frac{2M}{r}+\frac{\sqrt{r^2-a^2}}{r}+\frac{r^2}{\ell^2}-\frac{C_\omega}{r^{3\omega+1}}+\frac{Q^2}{r^2}.
\end{split}
\end{equation}

We must note that we require $r > a$ to avoid the emergence of imaginary structures. We consider the effects of quantum fluctuations when $r$ is significantly greater than $a$. We aim to thoroughly investigate its characteristics, focusing on thermodynamics, which is crucial to further investigation. Our study aims to elucidate the parameters that characterize the black hole. Here, $M$ represents the mass of the black hole, while $a$ relates to the quantum corrections that influence its properties. The symbol $\ell$ denotes the length scale pertinent to asymptotically AdS (Anti-de Sitter) spacetime. The parameter $C_\omega$ corresponds to the cosmological fluid surrounding the black hole, and $Q$ signifies the electric charge of the black hole.

\quad The mass and Hawking temperature of the quantum-corrected AdS-RN black hole surrounded by Kiselev spacetime are 
\begin{equation}\label{Unperturbed mass}
2M = \left(\sqrt{r_H ^2-a^2}- \frac{C_\omega}{r_H^{3 \omega }}+\frac{r_H^3}{\ell^2}+\frac{Q^2}{r_H}\right) \;\;;\;\; T_{H}=\frac{1}{4 \pi }\left(\frac{1}{\sqrt{r_H^2-a^2}}+ \frac{3 \omega C_\omega  }{ r_H^{3 \omega+2} }+\frac{3 r_H}{\ell^2}-\frac{Q^2}{r_H^3}\right) 
\end{equation}
It is clear that $r_H^2 > a^2$ for the real value of mass and temperature. The first law of thermodynamics effectively accommodates variations in the defining parameters, or hair, of the black hole, which include its area, cosmological constant, electric charge, quintessence parameter, and quantum correction parameter. For a comprehensive discussion on incorporating the quintessence parameter as a thermodynamic variable, our study further extends this framework to include the variability of the quantum correction parameter. 

\quad The first law of thermodynamics can be written as
\begin{equation}\label{eq3}
dM=TdS+VdP+\phi dQ+\mathcal{C}dC_\omega + \mathcal{A}da \ .
\end{equation}
Using Eq\eqref{eq3}, we can easily find the form of the other parameters as
\begin{equation*}
    \mathcal{C} = -\frac{1}{2 r_H^{3 \omega }} \;\;\;\;\;\;\;\;\;\;\;\;\;;\;\;\;\;\;\;\;\; \mathcal{A} = -\frac{a}{2  \sqrt{r_H^2-a^2}} \ .
\end{equation*}
The entropy of the quantum-corrected Schwarzschild black hole in Kiselev spacetime \cite{Sadeghi:2020ciy} aligns with the Hawking-Bekenstein entropy formula. Consequently, the entropy and pressure are 
\begin{equation}\label{eq4}
S = \pi r_H^2 \;\;\;\;\;\;\;\;\;\;;\;\;\;\;\;\;\;\;\;\; P = \frac{3}{8\pi\ell^2} \ .
\end{equation}
 The constraint on the value of horizon radius $r_H$ and quantum parameter $a$ as the regularization of the black hole singularity and is evident from the necessary condition as $S>\pi a^2$. 

\quad Recently, a universal thermodynamic extremality relation \cite{Goon:2019faz} was introduced by perturbing the metric that results in the perturbation of thermodynamic parameters. Based on that, the extremality relation is
\begin{equation}\label{Extremality Relation}
     \frac{\partial M_{ext}(\vec{\mathcal{Q}},\epsilon)}{\partial \epsilon} = \lim_{M \rightarrow M_{ext}} -T\left(\frac{\partial S(M,\vec{\mathcal{Q}},\epsilon)}{\partial \epsilon}\right)_{M,\vec{\mathcal{Q}}} \ ,
\end{equation}
where \( \epsilon \) represents the perturbation parameter. Here, \( M_{ext} \), \( T \), and \( S \) are the extremal mass bound (i.e., the mass at $T=0$), temperature, and entropy of the black hole post-correction, respectively, while \( \vec{\mathcal{Q}} \) represents the extensive thermodynamic variables of the black hole. The black hole solution under consideration offers a valuable opportunity to investigate whether any new equality emerges when other parameters are varied. We will examine the validity of the new extremality relation and compare it with the above one.
\section{The stability of extremal black hole with Weak Gravity Conjecture (WGC)}\label{Sec:Weak Gravity Conjucture}

 In this section, we will study the stability of extremal black holes with the help of WGC. The WGC is not merely a theoretical curiosity but a principle that may provide solutions to problems in cosmology and string theory, offer a framework for understanding black hole thermodynamics, and open new research avenues in quantum gravity phenomenology. The ongoing dialogue between theory and observation will continue to refine our understanding of the WGC and its implications across the universe. By connecting black hole thermodynamics with the WGC, we can bridge quantum and cosmic scales, leading to fresh perspectives on the implications and phenomenology associated with the conjecture. The WGC has intriguing connections to thermodynamics, particularly within black hole physics. In the framework of extra-dimensional theories, such as string theory, the WGC imposes constraints on the geometry of compactified dimensions. Even in more constrained geometries with fewer dimensions or symmetries, the WGC still implies the existence of particles that satisfy its mass-to-charge ratio requirements. In string theory, the WGC also plays a role in assessing the stability of non-supersymmetric (non-BPS) objects. In its mild form, the WGC suggests that black holes obtained from string theory compactified over certain manifolds lacking supersymmetry protection will ultimately decay. This decay is often linked to the presence of lighter particles in the theory’s spectrum. For instance, the authors of \cite{Marrani:2022jpt} examined the stability of non-BPS black holes and black strings within M-theory compactifications over Calabi-Yau threefolds. Using the WGC, they predicted the existence of minimal-volume cycles for a given homology class of Calabi-Yau threefolds, opening new avenues for mathematicians interested in manifold properties. In another study \cite{Mishra:2023ylz}, using three-parameter models, the authors demonstrated the existence of stable non-BPS extremal black holes, indicating the absence of lighter particles into which these black holes could decay. In the context of the AdS/CFT correspondence, the WGC translates into inequalities involving the dimensions and charges of operators, as well as central charges in the dual CFTs. However, these interpretations remain speculative and may not universally apply across all CFTs. In string theory, the WGC has been employed to argue for the existence of specific particles that affect the stability of extra dimensions and influence the mechanisms of supersymmetry breaking. A possible proof of the WGC within perturbative string theory connects the extremality bound of black holes to long-range forces calculated on the string worldsheet. The thermodynamics of black holes also provides a valuable framework for testing the WGC. The WGC thus acts as a bridge between quantum mechanics and large-scale cosmic phenomena, revealing how quantum-level interactions might manifest on cosmic scales and contributing to a deeper understanding of the holographic principle and the nature of spacetime. 

\quad WGC states that in any consistent theory of quantum gravity, there should be charged particles whose charge-to-mass ratio is greater than the extremal black hole\cite{Arkani-Hamed Nima}
\begin{equation}
    \frac{q}{m}\Big|_{charged\ particle} \geq \frac{Q}{M}\Big|_{ext- BH} \ .
\end{equation}
Another form of WGC states that the Charge mass ratio of extremal BH should be greater than $1$ in any Effective field theory(EFT) that admits UV completion. Achieve the goal with perturbative corrections and test the WGC. We will add the perturbation to the action, which is proportional to the cosmological constant. By modifying the action, the metric will also undergo corrections, affecting the thermodynamic quantities. We will derive these thermodynamic quantities from the perturbation parameter and check the universality relation as shown in \cite{Goon:2019faz}. 

\quad By making small perturbations in the metric, the metric is also changed as 
\begin{equation}
    g^{\text{total}}_{\mu \nu} = g_{\mu \nu} + \epsilon \;\mathrm{h}_{\mu \nu} \ ,
\end{equation}
where $\mathrm{h}_{\mu\nu}$ is the perturbed metric and $\epsilon$ denotes a pertubation parameter. The form of the perturbed metric reads 
\begin{eqnarray}
    ds_{\text{perturbed}}^2 = -\frac{r^2}{\ell^2} dt^2 -\frac{r^2}{\ell^2 f(r) }dr^2 \ .
\end{eqnarray}
The horizon shifted with the inclusion of perturbation parameter $\epsilon$. The thermodynamic quantities also get perturbed using the total metric, i.e., the metric obtained after perturbation. We will first compute the form of perturbed thermodynamically quantities. The perturbed mass of the black hole is 
\begin{equation}\label{Pertubed Mass}
    M(\epsilon) = \frac{\pi \ell^2 \left(\sqrt{S} \sqrt{S-\pi  a^2}+\pi  Q^2 \right)+S^2(1+\epsilon)}{2 \pi \ell^{3/2} \sqrt{S}}-\frac{1}{2} C_\omega \left(\frac{\pi}{S}\right) ^{3\omega/2 }  \ .
\end{equation}
Expanding into the power of $\epsilon$, it is easy to see 
\begin{equation}\label{M in terms of epsilon}
    M(\epsilon) = M +\frac{S^{3/2} }{2 \pi ^{3/2} \ell^2}  \epsilon + \mathcal{O}(\epsilon^2) \ ,
\end{equation}
where $M$ is unperturbed mass and its value is in Eq.\eqref{Unperturbed mass}. Using the Eq.\eqref{Pertubed Mass}and Eq.\eqref{Unperturbed mass}, the value of perturbation parameter $\epsilon$ is 
\begin{equation}\label{Epsilon}
    \epsilon =  \frac{\ell^2 \left(-\pi  \sqrt{S} \sqrt{S-\pi  a^2}+C_\omega \pi ^{3(\omega+1)/2 } S^{(1-3\omega)/2}+2 \pi ^{3/2} M \sqrt{S}-\pi ^2 Q^2\right)}{S^2}-1 \ .
\end{equation}
Now, it is easy to find the expression for $T \left(\frac{\partial S}{\partial \epsilon}\right)_{M,\Vec{\mathcal{Q}}}$, and its value is
\begin{equation*}
      -\frac{\pi  \ell^2 S^2 \left(3\omega C_\omega \pi ^{\frac{3 \omega+ }{2}}  \sqrt{S-\pi  a^2}-\pi  Q^2 \sqrt{S-\pi  a^2} S^{\frac{3 \omega-1 }{2}}+S^{\frac{3 \omega }{2}+1}\right)+3 (\epsilon +1) \sqrt{S-\pi  a^2} S^{\frac{3 \omega-7 }{2}}}{2 \pi ^{3/2} \ell^2 \left(\pi  \ell^2 \left(3 C_\omega \sqrt{S} \pi ^{\frac{3 \omega }{2}+\frac{1}{2}} \omega  \sqrt{S-\pi  a^2}-\pi  Q^2 \sqrt{S-\pi  a^2} S^{\frac{3 \omega }{2}}+S^{\frac{3 (\omega +1)}{2}}\right)+3 \sqrt{S-\pi  a^2} S^{\frac{3 \omega-1 }{2}}\right)} \ .
\end{equation*}
Now, using the Eq.\eqref{Epsilon} and Eq.\eqref{Unperturbed mass}, we can easily verify the universality relation \eqref{Extremality Relation}, i.e., 
\begin{equation}
      \left(\frac{\partial M_{\text{ext}}}{\partial \epsilon} \right)_{\vec{\mathcal{Q}}} = \, -T \left( \frac{\partial S}{\partial\epsilon} \right)_{M,\Vec{\mathcal{Q}}}   \ ,
\end{equation}
is satisfied. Now, it is natural to ask whether the other forms of universality conditions are satisfied, and we will first check the universality relation for pressure. For that, the expression for $V \left(\frac{\partial P}{\partial \epsilon}\right)_{M,\Vec{\mathcal{Q}}}$
\begin{equation*}
   \frac{32 P^2 S^{7/2}}{9 \sqrt{\pi } \left( \sqrt{S} \sqrt{S-\pi  a^2}-C_\omega \pi ^{\frac{3 \omega }{2}+\frac{1}{2}} S^{\frac{1}{2}-\frac{3 \omega }{2}}-2 \sqrt{S \pi } M +\pi  Q^2\right)}
\end{equation*}
Again, with the help of the unperturbed mass expression as in Eq.\eqref{Unperturbed mass}, we can easily verify another form of the universality relation\footnote{We have more hairs/parameters in the theory, they should also satisfy the universality condition and can be easily seen by computing them and using Eq.\eqref{M in terms of epsilon}, we have 
\begin{equation*}
    \Phi \left(\frac{\partial Q}{\partial \epsilon}\right) =   \mathcal{C} \left(\frac{\partial C_\omega}{\partial \epsilon}\right)=\mathcal{A} \left(\frac{\partial a}{\partial \epsilon}\right)= -\frac{S^{3/2}}{2 \pi ^{3/2} l^2 } =  \left(\frac{\partial M_{\text{ext}}}{\partial \epsilon} \right)_{\vec{\mathcal{Q}}} \ .
\end{equation*}}, also satisfied, i.e., 
\begin{equation}
      \left(\frac{\partial M_{\text{ext}}}{\partial \epsilon} \right)_{\vec{\mathcal{Q}}} = \lim_{M\rightarrow M_{ext}}\, -V \left( \frac{\partial P}{\partial\epsilon} \right)_{M,\Vec{\mathcal{Q}}}   \ .
\end{equation}

\begin{figure}[ht]
	\begin{center}
  \includegraphics[scale=0.70]{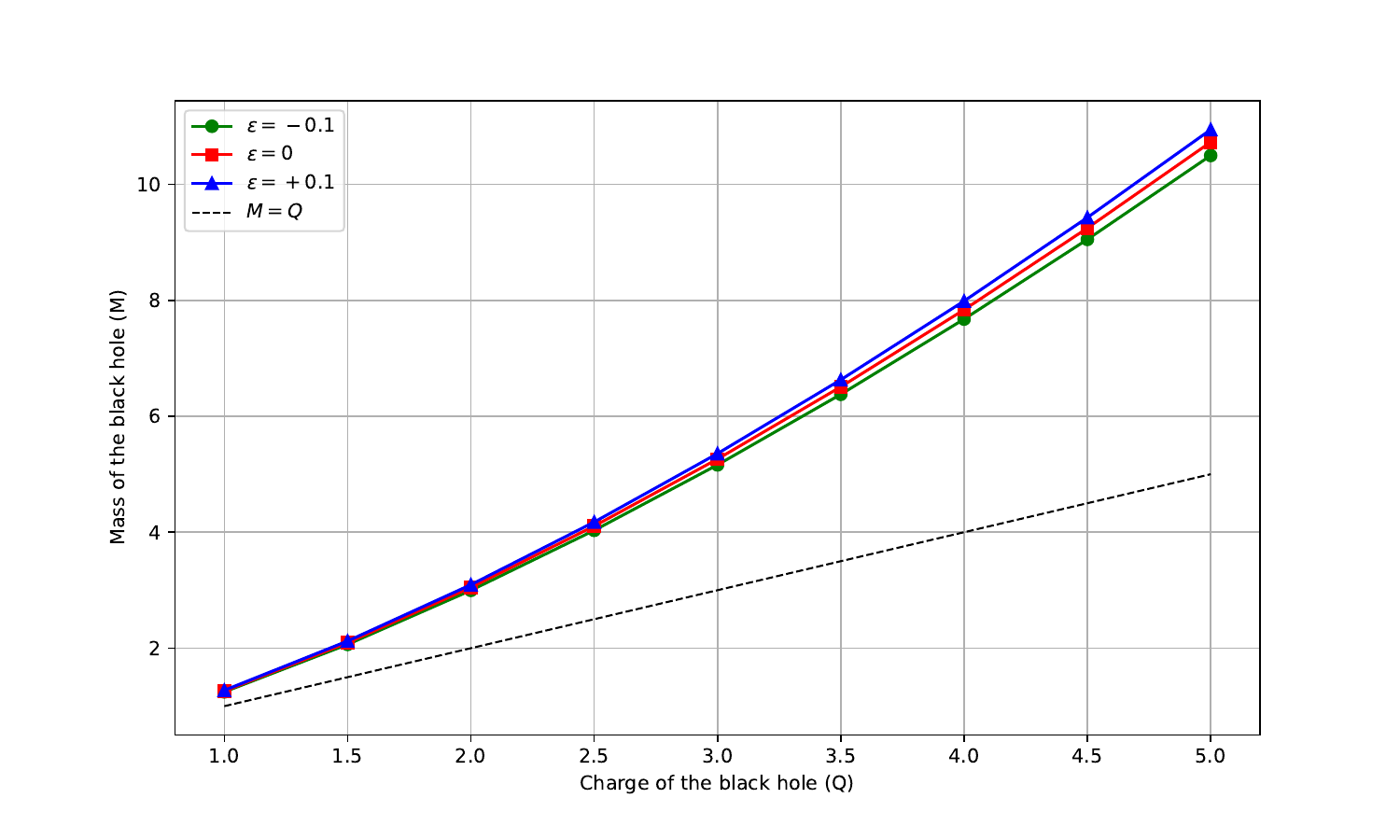}
	\end{center}
 \hspace{0.02Cm}
	\caption{Plot of mass of the black hole($M$) vs charge of the black hole($Q$) for different values of $\epsilon$ as well as $M=Q$ with dotted line.} 
	\label{fig:Plot $M$ vs $Q$}
\end{figure}

\quad Now, We analyze the effect of the perturbation on the extremal black hole and try to find whether we have a stable extremal black hole, i.e., a black hole with charge-to-mass ratio greater than $1$. To do our analysis, we have calculated the mass and temperature of a perturbed black hole in terms of perturbation parameter $\epsilon$. Using the fact that the temperature of an extremal black hole is zero gives us a relation between the charge and entropy of the black hole. It is impossible to invert the relation and obtain the entropy as an exact function of the charge of the black hole. Hence, we proceed numerically and obtain the entropy value for different values of black hole charges with some fixed values of other parameters. This we do for different values of $\epsilon$. Now, we use these values of charges and entropy to find the mass of the black hole. From Fig.\ref{fig:Plot $M$ vs $Q$}, we have obtained that we never find a situation in which the charge-to-mass ratio of these black holes is greater than $1$. We obtained that for the positive value of perturbation parameter $\epsilon$, the mass of the extremal black hole increases, causing a decrease in charge-to-mass ratio, and for the negative value of $\epsilon$, the mass of the extremal black hole decreases, causing an increase in mass to charge ratio. But we never encountered a situation in which the charge-to-mass ratio of the black hole was greater than unity, which is also expected from WGC as the decay to a black hole having a charge-to-mass ratio greater than unity is forbidden. Hence, we confirm WGC\cite{Arkani-Hamed Nima} in our system.

\quad The explanation we can draw from the above analysis is that 
the perturbation causes the black hole to emit or absorb the radiation, decreasing or increasing its mass. However, small perturbations are not able to make a black hole saturate its charge-to-mass ratio. Hence, these black holes are ultimately bound to decay, and we expect the existence of lighter particles into which the black hole can decay. The charge-to-mass ratio of these particles must be greater than that of the black hole.
\section{Charged Scalar Field}\label{Sec:Charged Scalar Field}
Our investigation focuses on the scattering of charged massive scalar fields in the context of an AdS-charged black hole in Kiselev spacetime. The dynamics of the charged massive scalar field, denoted by \(\Psi\), with mass \(\mu_{\text{s}}\) and charge \(q\), are governed by the equation of motion and can be written as 
\begin{equation}
    \frac{1}{\sqrt{-g}}(\partial_\mu - iqA_\mu)\left[\sqrt{-g}g^{\mu\nu}(\partial_\nu - iqA_\nu) \Psi\right] - \mu_{\text{s}}^2\Psi = 0 \ .
\end{equation}
Given the static and spherically symmetric nature of the spacetime, the scalar field can be expressed as
\begin{equation}
    \Psi(t,r,\theta,\phi) = e^{-i\omega t} R_{lm}(r)Y_{lm}(\theta,\phi) \ ,
\end{equation}
where \(R_{lm}(r)\) represents the radial function and \(Y_{lm}(\theta,\phi)\) are the spherical harmonics. By substituting this form into the equation of motion, we obtain the radial equation:
\begin{equation}
    \frac{1}{r^2} \frac{d}{dr} \left[r^2 f(r) \frac{dR_{lm}}{dr}\right] + \left[\frac{(\omega - \frac{qQ}{r})^2}{f(r)} - \frac{l(l+1)}{r^2} - \mu_{\text{s}}^2\right] R_{lm} = 0 \ ,
\end{equation}
and the angular equation is
\begin{equation}
    \left[\frac{1}{\sin\theta} \frac{\partial}{\partial\theta}\left( \sin\theta \frac{\partial}{\partial\theta}\right) + \frac{1}{\sin^2\theta} \frac{\partial^2}{\partial\phi^2}\right]Y_{lm} = -l(l+1)Y_{lm} \ .
\end{equation}
The angular solutions are well-established spherical harmonics, with the separation constant \(l(l+1)\), where \(l\) is a positive integer. As such, we focus on solving the radial equation. Introducing the tortoise coordinate \(r_*\), and defined by the relation $dr = f(r)\, dr_*$, the radial equation transforms into:
\begin{eqnarray}
    \frac{d^2R_{lm}}{dr_*^2} + \frac{2f(r)}{r} \frac{dR_{lm}}{dr_*} + \left[\left(\omega - \frac{qQ}{r}\right)^2 - f(r)\left(\frac{l(l+1)}{r^2} - \mu_{\text{s}}^2\right)\right]R_{lm} = 0 \ .
\end{eqnarray}
Near the event horizon (\(r = r_h\)), this simplifies to
\begin{equation}
    \frac{d^2R_{lm}}{dr_*^2} + \left(\omega - \frac{qQ}{r_h}\right)^2 R_{lm} = 0 \ ,
\end{equation}
which can be rewritten using the horizon's electric potential \(\phi_h\) as:
\begin{equation}
    \frac{d^2R_{lm}}{dr_*^2} + \left(\omega - q\phi_h\right)^2 R_{lm} = 0 \ .
\end{equation}
The solution near the horizon is:
\begin{equation}
    R_{lm}(r) \sim \exp[\pm i(\omega - q\phi_h)r_*] \ .
\end{equation}
Here, the positive sign corresponds to outgoing modes, while the negative sign corresponds to incoming modes. Since we are interested in the physically relevant incoming waves, we select the negative sign. Thus, the charged scalar field near the horizon takes the form:
\begin{equation}
    \Psi = \exp[-i(\omega - q\phi_h)r_*] Y_{lm}(\theta,\phi) e^{-i\omega t} \ .
\end{equation}
In the case of a non-rotating black hole, we restrict our analysis to a single wave mode \((l, m = 0)\). The evolution of the black hole's parameters during scattering can be deduced from the fluxes of the charged scalar field. The energy-momentum tensor is given by:
\begin{equation}
    T^\mu_\nu = \frac{1}{2} \mathcal{D}^\mu\Psi \partial_\nu\Psi^* + \frac{1}{2} \mathcal{D}^{*\mu}\Psi^*\mathcal{D}_\nu\Psi - \delta^\mu_\nu \left[\frac{1}{2} \mathcal{D}_\alpha \Psi \mathcal{D}^{*\alpha} \Psi^* - \frac{1}{2} \mu_{\text{s}} \Psi \Psi^*\right] \ ,
\end{equation}
where \(\mathcal{D} = \partial_\mu - iqA_\mu\). The energy flux through the horizon is derived as
\begin{equation}
    \frac{dE}{dt} = \int_{\text{H}} T^r_t \sqrt{-g} \, d\theta d\phi = \omega(\omega - q\phi_h)r_h^2 \ .
\end{equation}
The charge flux through the event horizon is similarly obtained as:
\begin{equation}
    \frac{dQ}{dt} = -\int_{\text{H}} j^r \sqrt{-g} \, d\theta d\phi = q(\omega - q\phi_h)r_h^2 \ ,
\end{equation}
where \(j^\mu\) is the electric current:
\begin{equation}
    j^\mu = -\frac{1}{2}iq(\Psi^*\mathcal{D}^\mu\Psi - \Psi \mathcal{D}^{*\mu}\Psi^*) \ .
\end{equation}
For wave modes with \(\omega > q \phi_h\), both energy and charge flow into the black hole. Conversely, when \(\omega < q \phi_h\), the fluxes are negative, indicating the extraction of energy and charge from the black hole a phenomenon known as black hole superradiance. Over an infinitesimal time interval \(dt\), the changes in mass and charge are given by
\begin{equation}\label{dMdEdQ}
    dM = dE = \omega(\omega - q\phi_h)r_h^2 \, dt, \quad dQ = q(\omega - q\phi_h)r_h^2 \, dt \ .
\end{equation}
In the above equation, fluxes will cause infinitesimal changes to the corresponding black hole properties over the small time interval, \(dt\). While the electric charge flux clearly leads to a variation in the black hole's charge, the energy flux is less straightforward, as it could represent changes in either the black hole's internal energy or its enthalpy. In this context, we will also establish similar relationships involving internal energy.

\section{Weak Cosmic Censorship Conjectures (WCCC)}\label{Sec:The Weak Cosmic Censorship Conjectures}
According to the WCCC, a black hole with a stable horizon will have its singularity obscured from an observer located at infinity by the horizon, thus preventing the formation of a naked singularity within the black hole's structure. In the context of RPS thermodynamics, this study examines the stability of the outer horizon while a charged scalar field scatters. During a short time interval $dt$, the initial state $f(\Vec{\mathcal{Q}})$ (where $\Vec{\mathcal{Q}}$ denotes the variable that is not constants) transitions to the final state 
$f(\Vec{\mathcal{Q}}+d \Vec{\mathcal{Q}})$ as it absorbs the scalar field. The existence of the horizon can be determined by analyzing solutions to $f(\Vec{\mathcal{Q}}+d \Vec{\mathcal{Q}}) = 0$, following the establishment of the outer horizon defined by $f(\Vec{\mathcal{Q}})=0$. The parameters associated with the quantum-corrected black holes in Kiselev spacetime are considered thermodynamic parameters to consider the influence of these quantities. $(M, Q, \ell, a, C_\omega, r_H)$ represents the black hole's inital state. $(M + dM, Q + dQ, \ell + d\ell, a + da, C_\omega+ dC_\omega, r_H + dr_H)$ represents the black hole's final state. Initially  $f(M, Q, \ell, a, C_\omega, r_H)$ fulfils the condition
\begin{equation*}
    f(M, Q, \ell, a, C_\omega, r_H) = 0 \ .
\end{equation*}
Also, by assuming that the black hole’s final state is still a black hole, it satisfies
\begin{equation*}
    f(M+dM, Q+dQ, \ell+d\ell, a+da, C_\omega+C_\omega, r_H+dr_H) = 0 \ .
\end{equation*}
We can relate these equations by assuming $\Vec{\mathcal{Q}}=(M, Q, \ell, a, C_\omega, r_H)$ as 
\begin{eqnarray}\label{ext T f1eqn}
 f\left(\Vec{\mathcal{Q}} + d\Vec{\mathcal{Q}} \right) &=& f(\Vec{\mathcal{Q}})+\frac{\partial f}{\partial M}dM+\frac{\partial f}{\partial Q}dQ+\frac{\partial f}{\partial \ell}d\ell+\frac{\partial f}{\partial a}da +\frac{\partial f}{\partial C_\omega}d C_\omega +\frac{\partial f}{\partial r}dr_{H} \ , \nonumber \\
 &=& f(\Vec{\mathcal{Q}})-\frac{2dM }{r_H} +\frac{2 Q dQ}{r_H^2}  -\frac{2  r_H^2 d\ell}{\ell^3} -\frac{a da}{ \sqrt{r_H^4-r_H^2a^2}} -\frac{ r_H^{-3 \omega }dC_\omega}{r_H} + 4 \pi  T dr_H  
\end{eqnarray}

With the condition that in small change in variables, i.e., $\Vec{\mathcal{Q}}$, the final state is still a black hole, and the above equation reduces to
\begin{equation}\label{Extended variation of drplus}
    dr_H =  \frac{a  \ell^2}{r_H \sqrt{r_H^2-a^2} \left(4 \pi  \ell^2 T-3 r_H\right)} da + \frac{ l^2 r_H^{-3 \omega -1}}{4 \pi  \ell^2 T-3 r_H} dC_\omega + \frac{2 r_H  \ell^2 \left(\omega -q \varphi_h \right)^2 }{ \left(4 \pi  \ell^2 T-3r_H\right)} dt \ .
    \end{equation}
Now, using the Eq.\eqref{Extended variation of drplus}, it is easy to see the change in entropy is 
\begin{eqnarray}\label{Extended variation of dS}
    dS =  \left[\frac{2\pi  a  \ell^2}{ \sqrt{r_H^2-a^2} \left(4 \pi  \ell^2 T-3 r_H\right)} da + \frac{2\pi \ell^2 r_H^{-3 \omega}}{4 \pi  \ell^2 T-3 r_H} dC_\omega + \frac{ 4\pi r_H^2  \ell^2 \left(\omega -q \varphi_h \right)^2 }{ \left(4 \pi  \ell^2 T-3r_H\right)} dt\right] \ .
\end{eqnarray}
The application of the second law of thermodynamics becomes ambiguous for an extremal black hole, where the temperature $T=0$. Classically, the second law asserts that the total entropy of a closed system should never decrease; however, for an extremal black hole, this law encounters challenges. The zero temperature and above equation lead to the possibility of scenarios where the entropy might decrease, which conflicts with the usual interpretation of the second law. This creates uncertainty regarding how the second law should be applied, as it is not straightforwardly upheld in the quantum-corrected framework where additional factors like quantum effects may further obscure the thermodynamic behavior of the black hole. Thus, the second law of thermodynamics for extremal black holes remains indefinite and subject to ongoing investigation.

\quad Let's focus on the near-extremal black hole and the situation when $T>3r_H/{4\pi \ell^2}$.  This condition relates to the black hole temperature and is derived from the thermodynamic properties of near-extremal black holes. It ensures that the temperature remains positive and prevents the breakdown of the second law of thermodynamics, i.e., entropy increase. In this case, the event horizon's area remains non-decreasing during the scattering of the complex scalar field, which is consistent with Hawking's area theorem. The theorem asserts that the area of a black hole's event horizon does not decrease during any classical process. As a result, the second law of thermodynamics holds, ensuring that the black hole's entropy increases.

\subsection*{Validity of WCCC}

\quad Now, the weak cosmic censorship conjecture can be tested by analyzing the function's behavior \( f(r) \). To ensure that the WCCC holds, one must evaluate the sign of the minimum value of \( f(r) \) at a particular radius, let's say \( r = r_{\text{min}} \). Specifically, if \( f(r) \) reaches its minimum at \( r_{\text{min}} \) and its value is $f_\text{min}$, and the value of the function at this point is negative (i.e., \( f(r_{\text{min}}) < 0 \)), this could indicate the formation of a naked singularity, thus violating the WCCC. In contrast, if \( f(r_{\text{min}}) \geq 0 \), the black hole horizon would still cover the singularity, and the conjecture remains valid. Therefore, determining whether the minimum value of \( f(r) \) is less than zero is crucial for assessing the robustness of the WCCC. 

\quad For an extremal black hole, the parameter \( f_\text{min}\) is equal to zero, while for a near-extremal black hole, \( f_\text{min} \) represents a small deviation. When the black hole absorbs the flux of a complex scalar field, the location of the minimum point shifts from \( r_{\text{min}} \) to \( r_{\text{min}} + dr_{\text{min}} \). Simultaneously, the black hole's parameters, originally \( (M, Q, \ell, a, C_\omega, r_H) \), will evolve to \( (M + dM, Q + dQ, \ell + d\ell, a + da, C_\omega+ dC_\omega, r_H + dr_H) \). Consequently, in the final state, the minimum value of the function \( f(r) \) is expressed in \eqref{ext T f1eqn}. Now by using the Eq.\eqref{ext T f1eqn} at $r=r_\text{min}$, we have 
\begin{eqnarray}\label{Fmin in delta}
 f\left(\Vec{\mathcal{Q}} + d\Vec{\mathcal{Q}} \right) &=& f_\text{min} -\frac{2 }{r_\text{min}}dM +\frac{2  Q}{r_\text{min}^2} dQ -\frac{2  r_\text{min}^2}{\ell^3}d\ell -\frac{a }{r_\text{min} \sqrt{r_\text{min}^2-a^2}} da -\frac{ r_\text{min}^{-2 \omega }}{r}dC_\omega \nonumber \\
 &=&f_\text{min} -\frac{2 T dS}{r_\text{min}} + \frac{2}{\ell^2}\left(\frac{r_H^3}{r_\text{min}}-r_\text{min}^2\right) d\ell - \frac{2Q}{r_\text{min}}\left(\frac{1}{r_H}-\frac{1}{r_\text{min}}\right) dQ \nonumber \\
 &&  +\frac{1}{r_\text{min}}\left(\frac{1}{r_H^{3\omega}}-\frac{1}{r_\text{min}^{3\omega}}\right) dC_{\omega} + \frac{a}{r_\text{min}}\left(\frac{1}{\sqrt{r_H^2-a^2}} - \frac{1}{\sqrt{r_\text{min}^2-a^2}}\right)da   \ .
\end{eqnarray}
In the case of an extremal black hole, the minimum radius \( r_{\text{min}} \) coincides with the event horizon radius \( r_H \), meaning that \( r_{\text{min}} = r_H \). Additionally, the temperature of the black hole, \( T \), is exactly zero at this extremal limit, signifying a state of zero thermal activity. By using these two conditions, it is easy to see 
\begin{eqnarray}
    f\left(\Vec{\mathcal{Q}} + d\Vec{\mathcal{Q}} \right) = 0 \ .
\end{eqnarray}
The scattering process in the context of an extremal black hole does not alter the minimum value of the function \( f(r) \). In an extremal black hole, this minimum value occurs at the event horizon and corresponds to a degenerate horizon, where the inner and outer horizons coincide, resulting in a zero-temperature black hole. Since the scattering process does not affect the minimum value of \( f(r) \), this implies that the horizon structure of the extremal black hole remains unchanged. Specifically, the degenerate horizon is preserved, meaning the black hole remains extremal even after the scattering event. This stability in the extremal configuration ensures that the black hole does not evolve into a non-extremal state. This result is crucial for the validity of the WCCC. So, due to this scattering, singularities are always hidden behind an event horizon, preventing naked singularities that would violate the predictability of spacetime. In conclusion, the fact that the scattering does not change the minimum value of \( f(r) \) proves that the extremal black hole retains its extremality, confirming that the horizon remains intact and the WCCC is valid.

\quad  For a near-extremal black hole, the condition of the metric function at the outer horizon, i.e.,  \( f'(r_H) \), is nearly zero, while the function value at the horizon satisfies \( f(r_H) = 0 \), and similarly, \( f'(r_{\text{min}}) = 0 \). To facilitate the calculation of Eq.\eqref{Fmin in delta}, we can assume that \( r_H = r_{\text{min}} + \delta \), where \( 0 < \delta \ll 1 \). By using this approximation, we can have 
\begin{eqnarray}
   f\left(\Vec{\mathcal{Q}} + d\Vec{\mathcal{Q}} \right) &=& f_\text{min}-\frac{2  T}{r_\text{min}} dS - \delta  \left(\frac{a \, da}{\left(r_\text{min}^2+a^2\right)^{3/2}}-\frac{6 r_\text{min}}{\ell^2} d\ell-\frac{2 Q \, dQ}{r_\text{min}^3}+ \frac{3 \omega \, dC_\omega}{  r_\text{min}^{3 \omega +2}}  \right) \nonumber \\
   && +\delta ^2 \left(-\frac{a \text{da} \left(a^2+2 r_{\text{min}}^2\right)}{2 r_{\text{min}} \left(r_{\text{min}}^2-a^2\right)^5/2}+\frac{3\omega (3 \omega +1)}{2} \frac{dC_\omega}{r_{\text{min}}^{3 \omega +3}}+\frac{6 d\ell}{\ell^2}-\frac{2 Q \, dQ }{r_{\text{min}}^4}\right) \ .
\end{eqnarray}
By setting the order of change in parameters of the order of $\delta$, i.e., $(dS, da, d\ell, dQ, dC_\omega) \sim \delta $, and only considering the terms up to the first order of $\delta$, we can easily write the above equation as 
\begin{eqnarray}
   f\left(\Vec{\mathcal{Q}} + d\Vec{\mathcal{Q}} \right) &=& f_\text{min}-\frac{2  T}{r_\text{min}} \delta + \mathcal{O}(\delta^2) <\, 0 \ ,
\end{eqnarray}
since $f_\text{min}$ is negative. As a result, the event horizon remains intact, ensuring that the black hole is not overcharged in its final state. This means that, despite any changes or perturbations, the charge-to-mass ratio of the black hole does not exceed the extremal limit, which would otherwise lead to the formation of a naked singularity. Since the event horizon continues, the singularity remains hidden from external observers, which aligns with the predictions of the WCCC. The conjecture holds true in the case of a near-extremal black hole, where the black hole is close to the extremal condition but not quite there. While examining the WCCC, it's crucial to account for the transfer of energy and charge through the fluxes of the scalar field over the time interval \(dt\). By introducing this time interval into our analysis, we assume that the energy and charge of the scalar field flow into the black hole in infinitesimally small amounts during \(dt\). The particles could transfer conserved quantities to overcharge the black hole beyond its extremal limit. To address this issue, it was suggested that the conserved quantities of the particles must be absorbed by the black hole in infinitesimally small increments. However, in the case of scalar field scattering, these small increments of energy and charge are inherently transferred over the infinitesimal time interval \(dt\). So, it ensures that the black hole is not overcharged, confirming that the WCCC is valid in this near-extremal scenario. 

\section{Conclusion}\label{Sec:Conclusion}

In this study, we explored and used the WGC in the context of quantum-corrected AdS-Reissner-Nordström (AdS-RN) black holes within Kiselev spacetime. We analyzed the extremality conditions of these black holes by incorporating corrections proportional to the cosmological constant within the framework of Kiselev spacetime. We examined the universal relation across all black hole parameters and confirmed that this universality holds for each hair of the black holes. 

\quad With the help of WGC, We explored the stability of extremal black holes in Kiselev spacetime. We observed that small perturbation causes extremal black holes to emit or receive radiations, ultimately decreasing or increasing the black hole mass for the given value of the charge. However, our analysis shows that small perturbations can not stabilize the black hole, i.e., the charge-to-mass ratio of the extremal black hole always remains lesser than $1$. We not only confirm the WGC but also, by using WGC, we predict that the black hole will decay by emitting charge particles whose charge-to-mass ratio is greater than that of the black hole. Hence, we predict these particles' existence in the theory's spectrum.

\quad We have also explored the WCCC in a quantum-corrected RN-AdS black hole within Kiselev spacetime, considering the scattering of a charged scalar field. We examined the changes in the black hole states due to the change in the black hole’s parameters, i.e., its charge $Q$, ads radius $\ell$, etc., over an infinitesimal time interval. We showed that if the condition \( T > \frac{3r_H}{4\pi \ell^2} \) is met, the second law of thermodynamics holds in the case of near-extremal black holes.
Finally, we showed the validity of WCCC  in both extremal and near-extremal black holes; hence, singularities formed within black holes must remain hidden from external observers, ensuring the preservation of the predictability of physical laws.

\quad There are several promising directions for future research that build upon the analysis of quantum corrections, including Exploring Quantum Corrections in Different Spacetime Geometries, Investigating Corrections to the Cosmic Censorship Conjectures, Studying Black Hole Stability under Quantum Corrections, Impact of Quantum Corrections on Black Hole Information Paradox, Phase Transitions and Critical Phenomena with Higher-Order Quantum Effects, Topological Analysis in Quantum-Corrected Spacetimes.

\quad \textbf{Note :} Indeed, while finalizing this manuscript, an investigation into the WGC validation within this model was conducted by Ref.\cite{Alipour:2024jgn}. In their study, the WGC was examined from the perspective of the photon sphere, focusing on how the properties of the photon sphere can be used to test and validate the conjecture in this context.


\subsection*{Acknowledgments}

AA and AM would like to express their deepest gratitude to Prasanta Tripathy for their invaluable guidance, constant support, and insightful feedback.  We are truly fortunate to have had the opportunity to learn a lot from him. PC is financially supported by Thailand NSRF via PMU-B under grant number PCB37G6600138.





\begin{thebibliography}{99}



\bibitem{Hawking:1975vcx}
S.~W.~Hawking, \emph{Particle Creation by Black Holes}, Commun. Math. Phys. \textbf{43} (1975), 199-220 [erratum: Commun. Math. Phys. \textbf{46} (1976), 206] \href{https://doi.org/10.1007/BF02345020}{doi:10.1007/BF02345020}.

\bibitem{Bardeen:1973gs}
J.~M.~Bardeen, B.~Carter and S.~W.~Hawking, \emph{The Four laws of black hole mechanics}, Commun. Math. Phys. \textbf{31} (1973), 161-170 \href{https://doi.org/10.1007/BF01645742}{doi:10.1007/BF01645742}.

\bibitem{PhysRevD.7.2333}
J.~D.~Bekenstein, \emph{Black Holes and Entropy}, Phys. Rev. D \textbf{7} (1973), 2333-2346 \href{https://doi.org/10.1103/PhysRevD.7.2333}{doi:10.1103/PhysRevD.7.2333}.

\bibitem{Hawking:1976de}
S.~W.~Hawking, \emph{Black Holes and Thermodynamics}, Phys. Rev. D \textbf{13} (1976), 191-197 \href{https://doi.org/10.1103/PhysRevD.13.191}{doi:10.1103/PhysRevD.13.191}.

\bibitem{Maldacena:1997re}
J.~M.~Maldacena, \emph{The Large N limit of superconformal field theories and supergravity}, Adv. Theor. Math. Phys. \textbf{2} (1998), 231-252 \href{https://doi.org/10.4310/ATMP.1998.v2.n2.a1}{doi:10.4310/ATMP.1998.v2.n2.a1} [\href{https://arxiv.org/abs/hep-th/9711200}{{\ttfamily arXiv:hep-th/9711200}}].

\bibitem{Palti Eran:2019}
E.~Palti, \emph{The Swampland: Introduction and Review}, Fortsch. Phys. \textbf{67} (2019) no.6, 1900037. \href{https://doi.org/10.1002/prop.201900037}{doi:10.1002/prop.201900037} [\href{https://arxiv.org/abs/1903.06239}{\ttfamily arXiv:1903.06239} [hep-th]].


\bibitem{vanBeest:2021lhn}
M.~van Beest, J.~Calder\'on-Infante, D.~Mirfendereski and I.~Valenzuela, \emph{Lectures on the Swampland Program in String Compactifications}, Phys. Rept. \textbf{989} (2022), 1-50 \href{https://doi.org/10.1016/j.physrep.2022.09.002}{doi:10.1016/j.physrep.2022.09.002} [\href{https://arxiv.org/abs/2102.01111}{{\ttfamily arXiv:2102.01111}}].

\bibitem{Arkani-Hamed Nima}
N.~Arkani-Hamed, L.~Motl, A.~Nicolis and C.~Vafa, \emph{The String landscape, black holes and gravity as the weakest force}, JHEP \textbf{06} (2007), 060. \href{https://doi.org/10.1088/1126-6708/2007/06/060}{doi:10.1088/1126-6708/2007/06/060} [\href{https://arxiv.org/abs/hep-th/0601001}{\ttfamily hep-th/0601001}].

\bibitem{Harlow}
D.~Harlow, B.~Heidenreich, M.~Reece and T.~Rudelius, \emph{Weak gravity conjecture}, Rev. Mod. Phys. \textbf{95} (2023) no.3, 3. \href{https://doi.org/10.1103/RevModPhys.95.035003}{doi:10.1103/RevModPhys.95.035003} [\href{https://arxiv.org/abs/2201.08380}{\ttfamily arXiv:2201.08380}].

\bibitem{Vafa:2005ui}
C.~Vafa, \emph{The String landscape and the swampland}, [\href{https://arxiv.org/abs/hep-th/0509212}{{\ttfamily arXiv:hep-th/0509212}}].

\bibitem{Ooguri:2006in}
H.~Ooguri and C.~Vafa, \emph{On the Geometry of the String Landscape and the Swampland}, Nucl. Phys. B \textbf{766} (2007), 21-33 \href{https://doi.org/10.1016/j.nuclphysb.2006.10.033}{doi:10.1016/j.nuclphysb.2006.10.033} [\href{https://arxiv.org/abs/hep-th/0605264}{{\ttfamily arXiv:hep-th/0605264}}].

\bibitem{Arkani-Hamed:2006emk}
N.~Arkani-Hamed, L.~Motl, A.~Nicolis and C.~Vafa, \emph{The String landscape, black holes and gravity as the weakest force}, JHEP \textbf{06} (2007), 060 \href{https://doi.org/10.1088/1126-6708/2007/06/060}{doi:10.1088/1126-6708/2007/06/060} [\href{https://arxiv.org/abs/hep-th/0601001}{{\ttfamily arXiv:hep-th/0601001}}].

\bibitem{Heidenreich:2016aqi}
B.~Heidenreich, M.~Reece and T.~Rudelius, \emph{Evidence for a sublattice weak gravity conjecture}, JHEP \textbf{08} (2017), 025 \href{https://doi.org/10.1007/JHEP08(2017)025}{doi:10.1007/JHEP08(2017)025} [\href{https://arxiv.org/abs/1606.08437}{{\ttfamily arXiv:1606.08437}}].

\bibitem{Palti:2017elp}
E.~Palti, \emph{The Weak Gravity Conjecture and Scalar Fields}, JHEP \textbf{08} (2017), 034 \href{https://doi.org/10.1007/JHEP08(2017)034}{doi:10.1007/JHEP08(2017)034} [\href{https://arxiv.org/abs/1705.04328}{{\ttfamily arXiv:1705.04328}}].

\bibitem{Odintsov:2020zkl}
S.~D.~Odintsov and V.~K.~Oikonomou, \emph{Swampland implications of GW170817-compatible Einstein-Gauss-Bonnet gravity}, Phys. Lett. B \textbf{805} (2020), 135437 \href{https://doi.org/10.1016/j.physletb.2020.135437}{doi:10.1016/j.physletb.2020.135437} [\href{https://arxiv.org/abs/2004.00479}{{\ttfamily arXiv:2004.00479}}].




\bibitem{Liu:2021diz}
Y.~Liu, \emph{Higgs inflation and its extensions and the further refining dS swampland conjecture}, \href{https://doi.org/10.1140/epjc/s10052-021-09940-w}{\emph{Eur. Phys. J. C} {\bfseries 81} (2021) 1122} [\href{https://arxiv.org/abs/2112.14571}{{\ttfamily arXiv:2112.14571}}].

\bibitem{Liu:2022myp}
Y.~Liu, \emph{Higgs inflation and scalar weak gravity conjecture}, \href{https://doi.org/10.1140/epjc/s10052-022-10993-8}{\emph{Eur. Phys. J. C} {\bfseries 82} (2022) 1052} [\href{https://arxiv.org/abs/2205.00576}{{\ttfamily arXiv:2205.00576}}].

\bibitem{g}
S.~Noori Gashti and J.~Sadeghi, \emph{Refined swampland conjecture in warm vector hybrid inflationary scenario}, \href{https://doi.org/10.1140/epjp/s13360-022-02423-7}{\emph{The European Physical Journal Plus} {\bfseries 137} (2022) 1-13}.

\bibitem{Alipour:2023css}
M.~R.~Alipour, J.~Sadeghi and M.~Shokri, \emph{WGC and WCC for charged black holes with quintessence and cloud of strings}, \href{https://doi.org/10.1140/epjc/s10052-023-11811-5}{\emph{Eur. Phys. J. C} {\bfseries 83} (2023) 640} [\href{https://arxiv.org/abs/2307.09654}{{\ttfamily arXiv:2307.09654}}].

\bibitem{Alipour:2023yiz}
M.~R.~Alipour, J.~Sadeghi and M.~Shokri, \emph{WGC and WCCC of black holes with quintessence and cloud strings in RPS space}, \href{https://doi.org/10.1016/j.nuclphysb.2023.116184}{\emph{Nucl. Phys. B} {\bfseries 990} (2023) 116184} [\href{https://arxiv.org/abs/2303.02487}{{\ttfamily arXiv:2303.02487}}].

\bibitem{Sadeghi:2023cui}
J.~Sadeghi, M.~R.~Alipour and S.~Noori Gashti, \emph{Emerging WGC from the Dirac particle around black holes}, \href{https://doi.org/10.1142/S0217732323501225}{\emph{Mod. Phys. Lett. A} {\bfseries 38} (2023) 2350122}.

\bibitem{Schoneberg:2023lun}
N.~Sch\"oneberg, L.~Vacher, J.~D.~F.~Dias, M.~M.~C.~D.~Carvalho and C.~J.~A.~P.~Martins, \emph{News from the Swampland — constraining string theory with astrophysics and cosmology}, \href{https://doi.org/10.1088/1475-7516/2023/10/039}{\emph{JCAP} {\bfseries 10} (2023) 039} [\href{https://arxiv.org/abs/2307.15060}{{\ttfamily arXiv:2307.15060}}].

\bibitem{Kadota:2019dol}
K.~Kadota, C.~S.~Shin, T.~Terada and G.~Tumurtushaa, \emph{Trans-Planckian censorship and single-field inflaton potential}, \href{https://doi.org/10.1088/1475-7516/2020/01/008}{\emph{JCAP} {\bfseries 01} (2020) 008} [\href{https://arxiv.org/abs/1910.09460}{{\ttfamily arXiv:1910.09460}}].

\bibitem{Oikonomou:2020oex}
V.~K.~Oikonomou, \emph{Rescaled Einstein-Hilbert Gravity from $f(R)$ Gravity: Inflation, Dark Energy and the Swampland Criteria}, \href{https://doi.org/10.1103/PhysRevD.103.124028}{\emph{Phys. Rev. D} {\bfseries 103} (2021) 124028} [\href{https://arxiv.org/abs/2012.01312}{{\ttfamily arXiv:2012.01312}}].

\bibitem{Sadeghi:2023cxh}
J.~Sadeghi, B.~Pourhassan, S.~Noori Gashti, \.I.~Sakall\i{} and M.~R.~Alipour, \emph{de Sitter swampland conjecture in string field inflation}, \href{https://doi.org/10.1140/epjc/s10052-023-11822-2}{\emph{Eur. Phys. J. C} {\bfseries 83} (2023) 635} [\href{https://arxiv.org/abs/2303.04551}{{\ttfamily arXiv:2303.04551}}].

\bibitem{NooriGashti:2022xmf}
S.~Noori Gashti, J.~Sadeghi, S.~Upadhyay and M.~R.~Alipour, \emph{Swampland dS conjecture in mimetic $f(R, T)$ gravity}, \href{https://doi.org/10.1088/1572-9494/ac7a1f}{\emph{Commun. Theor. Phys.} {\bfseries 74} (2022) 085402} [\href{https://arxiv.org/abs/2208.12229}{{\ttfamily arXiv:2208.12229}}].

\bibitem{Das:2019hto}
S.~Das, \emph{Distance, de Sitter and Trans-Planckian Censorship conjectures: the status quo of Warm Inflation}, \href{https://doi.org/10.1016/j.dark.2019.100432}{\emph{Phys. Dark Univ.} {\bfseries 27} (2020) 100432} [\href{https://arxiv.org/abs/1910.02147}{{\ttfamily arXiv:1910.02147}}].

\bibitem{Yuennan:2022zml}
J.~Yuennan and P.~Channuie, \emph{Further Refining Swampland Conjecture on Inflation in General Scalar-Tensor Theories of Gravity}, \href{https://doi.org/10.1002/prop.202200024}{\emph{Fortsch. Phys.} {\bfseries 70} (2022) 2200024} [\href{https://arxiv.org/abs/2202.02690}{{\ttfamily arXiv:2202.02690}}].

\bibitem{Bedroya:2019snp}
A.~Bedroya and C.~Vafa, \emph{Trans-Planckian Censorship and the Swampland}, \href{https://doi.org/10.1007/JHEP09(2020)123}{\emph{JHEP} {\bfseries 09} (2020) 123} [\href{https://arxiv.org/abs/1909.11063}{{\ttfamily arXiv:1909.11063}}].

\bibitem{Sadeghi:2023cpd}
J.~Sadeghi, S.~Noori Gashti, M.~R.~Alipour and M.~A.~S.~Afshar, \emph{Can black holes cause cosmic expansion?}, [\href{https://arxiv.org/abs/2305.12545}{{\ttfamily arXiv:2305.12545}}].


\bibitem{v}
A.~Mohammadi, T.~Golanbari and J.~Enayati, \emph{{Brane inflation and Trans-Planckian censorship conjecture}}, \href{https://doi.org/10.1103/PhysRevD.104.123515}{\emph{Phys. Rev. D} {\bfseries 104} (2021) 123515}.

\bibitem{w}
J.~Sadeghi, M.~R.~Alipour and S.~Noori Gashti, \emph{{Scalar Weak Gravity Conjecture in Super Yang-Mills Inflationary Model}}, \href{https://doi.org/10.3390/universe8120621}{\emph{Universe} {\bfseries 8} (2022) 621}.

\bibitem{x}
R.~Kallosh, et al., \emph{{dS Vacua and the Swampland}}, \href{https://doi.org/10.1007/JHEP03(2019)013}{\emph{J. High Energy Phys.} {\bfseries 2019} (2019) 1-18}.

\bibitem{Guleryuz:2021zik}
O.~Guleryuz, \emph{{On the Trans-Planckian Censorship Conjecture and the generalized non-minimal coupling}}, \href{https://doi.org/10.1088/1475-7516/2021/11/043}{\emph{JCAP} {\bfseries 11} (2021) no.11, 043} [\href{https://arxiv.org/abs/2105.10571}{{\ttfamily 2105.10571}}].

\bibitem{Osses:2021snt}
C.~Osses, N.~Videla and G.~Panotopoulos, \emph{{Reheating in small-field inflation on the brane: The Swampland Criteria and observational constraints in light of the PLANCK 2018 results}}, \href{https://doi.org/10.1140/epjc/s10052-021-09283-6}{\emph{Eur. Phys. J. C} {\bfseries 81} (2021) no.6, 485} [\href{https://arxiv.org/abs/2101.08882}{{\ttfamily 2101.08882}}].

\bibitem{aa}
J.~Sadeghi, S.~Noori Gashti and M.~R.~Alipour, \emph{{Notes on further refining de Sitter swampland conjecture with inflationary models}}, \href{https://doi.org/10.1016/j.cjp.2022.03.007}{\emph{Chinese J. Phys.} {\bfseries 79} (2022) 490-502}.

\bibitem{Brahma:2019unn}
S.~Brahma, \emph{{Trans-Planckian censorship, inflation and excited initial states for perturbations}}, \href{https://doi.org/10.1103/PhysRevD.101.023526}{\emph{Phys. Rev. D} {\bfseries 101} (2020) no.2, 023526} [\href{https://arxiv.org/abs/1910.04741}{{\ttfamily 1910.04741}}].

\bibitem{Brandenberger:2021pzy}
R.~Brandenberger, \emph{{Trans-Planckian Censorship Conjecture and Early Universe Cosmology}}, \href{https://doi.org/10.31526/lhep.2021.198}{\emph{LHEP} {\bfseries 2021} (2021), 198} [\href{https://arxiv.org/abs/2102.09641}{{\ttfamily 2102.09641}}].

\bibitem{Sadeghi:2022tzd}
J.~Sadeghi, S.~Noori Gashti and F.~Darabi, \emph{{Swampland conjectures in hybrid metric-Palatini gravity}}, \href{https://doi.org/10.1016/j.dark.2022.101090}{\emph{Phys. Dark Univ.} {\bfseries 37} (2022), 101090} [\href{https://arxiv.org/abs/2207.09793}{{\ttfamily 2207.09793}}].

\bibitem{Geng:2019bnn}
H.~Geng, S.~Grieninger and A.~Karch, \emph{{Entropy, Entanglement and Swampland Bounds in DS/dS}}, \href{https://doi.org/10.1007/JHEP06(2019)105}{\emph{JHEP} {\bfseries 06} (2019), 105} [\href{https://arxiv.org/abs/1904.02170}{{\ttfamily 1904.02170}}].

\bibitem{Gashti:2022hey}
S.~N.~Gashti, J.~Sadeghi and B.~Pourhassan, \emph{{Pleasant behavior of swampland conjectures in the face of specific inflationary models}}, \href{https://doi.org/10.1016/j.astropartphys.2022.102703}{\emph{Astropart. Phys.} {\bfseries 139} (2022), 102703} [\href{https://arxiv.org/abs/2202.06381}{{\ttfamily 2202.06381}}].

\bibitem{gg}
J.~Sadeghi, E.~Naghd Mezerji and S.~Noori Gashti, \emph{{Study of some cosmological parameters in logarithmic corrected $f(R)$ gravitational model with swampland conjectures}}, \href{https://doi.org/10.1142/S0217732321500274}{\emph{Mod. Phys. Lett. A} {\bfseries 36} (2021) 2150027}.

\bibitem{hh}
J.~Sadeghi, S.~Noori Gashti and E.~Naghd Mezerji, \emph{{The investigation of universal relation between corrections to entropy and extremality bounds with verification WGC}}, \href{https://doi.org/10.1016/j.dark.2020.100626}{\emph{Phys. Dark Univ.} {\bfseries 30} (2020), 100626}.

\bibitem{Agrawal:2018own}
P.~Agrawal, G.~Obied, P.~J.~Steinhardt and C.~Vafa, \emph{{On the Cosmological Implications of the String Swampland}}, \href{https://doi.org/10.1016/j.physletb.2018.07.040}{\emph{Phys. Lett. B} {\bfseries 784} (2018), 271-276} [\href{https://arxiv.org/abs/1806.09718}{{\ttfamily 1806.09718}}].

\bibitem{kk}
J.~Sadeghi and S.~Noori Gashti, \emph{{Investigating the logarithmic form of $f(R)$ gravity model from brane perspective and swampland criteria}}, \href{https://doi.org/10.1007/s12043-021-2044-y}{\emph{Pramana} {\bfseries 95} (2021) 1-8}.

\bibitem{ll}
U.~Kumar Sharma, \emph{{Reconstruction of quintessence field for the THDE with swampland correspondence in $f(R,T)$ gravity}}, \href{https://doi.org/10.1142/S0219887821500310}{\emph{Int. J. Geom. Meth. Mod. Phys.} {\bfseries 18} (2021) 2150031}.

\bibitem{mm}
J.~Sadeghi, M.~R.~Alipour and S.~Noori Gashti, \emph{{Strong cosmic censorship in light of weak gravity conjecture for charged black holes}}, \href{https://doi.org/10.1007/JHEP02(2023)017}{\emph{J. High Energy Phys.} {\bfseries 2023} (2023) 1-14}.

\bibitem{Odintsov:2017hbk}
S.~D.~Odintsov, V.~K.~Oikonomou and L.~Sebastiani, \emph{{Unification of Constant-roll Inflation and Dark Energy with Logarithmic $R^2$-corrected and Exponential $F(R)$ Gravity}}, \href{https://doi.org/10.1016/j.nuclphysb.2017.08.018}{\emph{Nucl. Phys. B} {\bfseries 923} (2017) 608-632} [\href{https://arxiv.org/abs/1708.08346}{{\ttfamily 1708.08346}}].

\bibitem{Shokri:2021zqw}
M.~Shokri, J.~Sadeghi and S.~N.~Gashti, \emph{{Quintessential constant-roll inflation}}, \href{https://doi.org/10.1016/j.dark.2021.100923}{\emph{Phys. Dark Univ.} {\bfseries 35} (2022) 100923} [\href{https://arxiv.org/abs/2107.04756}{{\ttfamily 2107.04756}}].

\bibitem{Sadeghi:2021plz}
J.~Sadeghi, B.~Pourhassan, S.~Noori Gashti and S.~Upadhyay, \emph{{Swampland conjecture and inflation model from brane perspective}}, \href{https://doi.org/10.1088/1402-4896/ac39bc}{\emph{Phys. Scripta} {\bfseries 96} (2021) 125317} [\href{https://arxiv.org/abs/2111.15477}{{\ttfamily 2111.15477}}].

\bibitem{Shokri:2021hjs}
M.~Shokri, J.~Sadeghi, R.~Herrera and S.~Noori Gashti, \emph{{Warm inflation with bulk viscous pressure for different solutions of an anisotropic universe}}, \href{https://doi.org/10.1088/1402-4896/ad6b52}{\emph{Phys. Scripta} {\bfseries 99} (2024) 095010} [\href{https://arxiv.org/abs/2112.12309}{{\ttfamily 2112.12309}}].

\bibitem{Sadeghi:2020xtc}
J.~Sadeghi, B.~Pourhassan, S.~Noori Gashti, S.~Upadhyay and E.~N.~Mezerji, \emph{{The emergence of universal relations in the AdS black holes thermodynamics}}, \href{https://doi.org/10.1088/1402-4896/acb40b}{\emph{Phys. Scripta} {\bfseries 98} (2023) 025305} [\href{https://arxiv.org/abs/2011.14366}{{\ttfamily 2011.14366}}].

\bibitem{Yuennan:2022vdd}
J.~Yuennan and P.~Channuie, \emph{{Composite inflation and further refining dS swampland conjecture}}, \href{https://doi.org/10.1016/j.nuclphysb.2022.116033}{\emph{Nucl. Phys. B} {\bfseries 986} (2023) 116033} [\href{https://arxiv.org/abs/2208.09842}{{\ttfamily 2208.09842}}].

\bibitem{uu}
S. Noori Gashti and J. Sadeghi, \emph{{Refined swampland conjecture in warm vector hybrid inflationary scenario}}, \href{https://doi.org/10.1140/epjp/s13360-022-01679-9}{\emph{The European Physical Journal Plus} {\bfseries 137} (2022) 1-13}.

\bibitem{Kinney:2018kew}
W.~H.~Kinney, \emph{{Eternal Inflation and the Refined Swampland Conjecture}}, \href{https://doi.org/10.1103/PhysRevLett.122.081302}{\emph{Phys. Rev. Lett.} {\bfseries 122} (2019) 081302} [\href{https://arxiv.org/abs/1811.11698}{{\ttfamily 1811.11698}}].

\bibitem{Kinney:2021hje}
W.~H.~Kinney, \emph{{The Swampland Conjecture Bound Conjecture}}, [\href{https://arxiv.org/abs/2103.16583}{{\ttfamily 2103.16583}}].

\bibitem{Yu:2018eqq}
T.~Y.~Yu and W.~Y.~Wen, \emph{{Cosmic censorship and Weak Gravity Conjecture in the Einstein–Maxwell-dilaton theory}}, \href{https://doi.org/10.1016/j.physletb.2018.04.060}{\emph{Phys. Lett. B} {\bfseries 781} (2018) 713-718} [\href{https://arxiv.org/abs/1803.07916}{{\ttfamily 1803.07916}}].

\bibitem{NooriGashti:2021nox}
S.~Noori Gashti, \emph{{Two-field inflationary model and swampland de Sitter conjecture}}, \href{https://doi.org/10.22128/jhap.2021.452.1002}{\emph{JHAP} {\bfseries 2} (2022) no.1, 13-24} [\href{https://arxiv.org/abs/2111.06421}{{\ttfamily 2111.06421}}].

\bibitem{Sadeghi:2022wgx}
J.~Sadeghi, M.~Shokri, M.~R.~Alipour and S.~Noori Gashti, \emph{{Weak gravity conjecture from conformal field theory: a challenge from hyperscaling violating and Kerr-Newman-AdS black holes}}, \href{https://doi.org/10.1088/1674-1137/ac957b}{\emph{Chin. Phys. C} {\bfseries 47} (2023) no.1, 015103} [\href{https://arxiv.org/abs/2203.03378}{{\ttfamily 2203.03378}}].

\bibitem{Sadeghi:2022xcr}
J.~Sadeghi, B.~Pourhassan, S.~Noori Gashti and S.~Upadhyay, \emph{{Weak gravity conjecture, black branes and violations of universal thermodynamics relation}}, \href{https://doi.org/10.1016/j.aop.2022.169168}{\emph{Annals Phys.} {\bfseries 447} (2022) 169168} [\href{https://arxiv.org/abs/2201.04071}{{\ttfamily 2201.04071}}].

\bibitem{Kolb:2021nob}
E.~W.~Kolb, A.~J.~Long and E.~McDonough, \emph{{Gravitino Swampland Conjecture}}, \href{https://doi.org/10.1103/PhysRevLett.127.131603}{\emph{Phys. Rev. Lett.} {\bfseries 127} (2021) no.13, 131603} [\href{https://arxiv.org/abs/2106.03654}{{\ttfamily 2106.03654}}].



\bibitem{Sadeghi:2021zjb}
J.~Sadeghi and S.~Noori Gashti, \emph{{Investigating the validity of the weak cosmic censorship conjecture for the three charged black holes}}, \href{https://doi.org/10.1016/j.physletb.2021.136542}{\emph{Phys. Lett. B} {\bfseries 818} (2021) 136542} [\href{https://arxiv.org/abs/2106.14738}{{\ttfamily 2106.14738}}].


\bibitem{Cho:2023koe}
M.~Cho, S.~Choi, K.~H.~Lee and J.~Song, \emph{{Supersymmetric Cardy formula and the Weak Gravity Conjecture in AdS/CFT}}, \href{https://doi.org/10.1007/JHEP11(2023)118}{\emph{JHEP} \textbf{11} (2023) 118} [\href{https://arxiv.org/abs/2308.01717}{{\ttfamily 2308.01717}}].

\bibitem{Nakayama:2015hga}
Y.~Nakayama and Y.~Nomura, \emph{{Weak gravity conjecture in the AdS/CFT correspondence}}, \href{https://doi.org/10.1103/PhysRevD.92.126006}{\emph{Phys. Rev. D} \textbf{92} (2015) no.12, 126006} [\href{https://arxiv.org/abs/1509.01647}{{\ttfamily 1509.01647}}].

\bibitem{Aalsma:2020duv}
L.~Aalsma, A.~Cole, G.~J.~Loges and G.~Shiu, \emph{{A New Spin on the Weak Gravity Conjecture}}, \href{https://doi.org/10.1007/JHEP03(2021)085}{\emph{JHEP} \textbf{03} (2021) 085} [\href{https://arxiv.org/abs/2011.05337}{{\ttfamily 2011.05337}}].

\bibitem{Heidenreich:2021yda}
B.~Heidenreich, M.~Reece and T.~Rudelius, \emph{{The Weak Gravity Conjecture and axion strings}}, \href{https://doi.org/10.1007/JHEP11(2021)004}{\emph{JHEP} \textbf{11} (2021) 004} [\href{https://arxiv.org/abs/2108.11383}{{\ttfamily 2108.11383}}].

\bibitem{Loges:2019jzs}
G.~J.~Loges, T.~Noumi and G.~Shiu, \emph{{Thermodynamics of 4D Dilatonic Black Holes and the Weak Gravity Conjecture}}, \href{https://doi.org/10.1103/PhysRevD.102.046010}{\emph{Phys. Rev. D} \textbf{102} (2020) no.4, 046010} [\href{https://arxiv.org/abs/1909.01352}{{\ttfamily 1909.01352}}].

\bibitem{JiZh20}
J.~Jiang and M.~Zhang, \emph{{Weak cosmic censorship conjecture in Einstein-Maxwell gravity with scalar hair}}, \href{https://doi.org/10.1140/epjc/s10052-020-7822-6}{\emph{Eur. Phys. J. C} \textbf{80} (2020) 196}.




\bibitem{GMZZ18}
B.~Ge, Y.~Mo, S.~Zhao and J.~Zheng, \emph{{Higher-dimensional charged black holes cannot be over-charged by gedanken experiments}}, \href{https://doi.org/10.1016/j.physletb.2018.06.031}{\emph{Phys. Lett. B} \textbf{783} (2018) 440} [\href{https://arxiv.org/abs/1712.07342}{{\ttfamily 1712.07342}}].

\bibitem{NiCL19}
B.~Ning, B.~Chen and F.~L.~Lin, \emph{{Gedanken Experiments to Destroy a BTZ Black Hole}}, \href{https://doi.org/10.1103/PhysRevD.100.044043}{\emph{Phys. Rev. D} \textbf{100} (2019) 044043} [\href{https://arxiv.org/abs/1902.00949}{{\ttfamily 1902.00949}}].

\bibitem{DuSe13}
K.~D\"uztas and I.~Semiz, \emph{{Cosmic Censorship, Black Holes and Integer-spin Test Fields}}, \href{https://doi.org/10.1103/PhysRevD.88.064043}{\emph{Phys. Rev. D} {\bf 88} (2013) 064043} [\href{https://arxiv.org/abs/1307.1481}{{\ttfamily 1307.1481}}].

\bibitem{SeDu15}
I.~Semiz and K.~D\"uztas, \emph{{Weak Cosmic Censorship, Superradiance and Quantum Particle Creation}}, \href{https://doi.org/10.1103/PhysRevD.92.104021}{\emph{Phys. Rev. D} {\bf 92} (2015) 104021} [\href{https://arxiv.org/abs/1507.03744}{{\ttfamily 1507.03744}}].

\bibitem{Duzt15}
K.~D\"uztas, \emph{{Stability of event horizons against neutrino flux: The classical picture}}, \href{https://doi.org/10.1088/0264-9381/32/7/075003}{\emph{Class. Quant. Grav.} {\bf 32} (2015) 075003} [\href{https://arxiv.org/abs/1408.1735}{{\ttfamily 1408.1735}}].




\bibitem{Kiselev:2002dx}
V.~V.~Kiselev, \emph{{Quintessence and black holes}}, \href{https://10.1088/0264-9381/20/6/310}{\emph{Class. Quant. Grav. } {\bfseries 20} (2003) 1187-1198} [\href{https://arxiv.org/abs/gr-qc/0210040}{{\ttfamily gr-qc/0210040 [gr-qc]}}].

\bibitem{Visser:2019brz}
M.~Visser, \emph{{The Kiselev black hole is neither perfect fluid, nor is it quintessence}}, \href{https://doi.org/10.1088/1361-6382/ab60b8}{\emph{Class. Quant. Grav.} {\bfseries 37} (2020) 045001} [\href{https://arxiv.org/abs/1908.11058}{{\ttfamily 1908.11058}}].


\bibitem{Kazakov:1993ha}
D.~I.~Kazakov and S.~N.~Solodukhin, \emph{{On Quantum deformation of the Schwarzschild solution}}, \href{https://doi.org/10.1016/S0550-3213(94)80045-6}{\emph{Nucl. Phys. B} {\bfseries 429} (1994) 153-176} [\href{https://arxiv.org/abs/hep-th/9310150}{{\ttfamily hep-th/9310150}}].


\bibitem{MoraisGraca:2021ife}
J.~P.~Morais Gra\c{c}a, E.~Folco Capossoli, H.~Boschi-Filho and I.~P.~Lobo, \emph{{Joule-Thomson expansion for quantum corrected AdS-Reissner-N\"ordstrom black holes in a Kiselev spacetime}}, \href{https://doi.org/10.1103/PhysRevD.107.024045}{\emph{Phys. Rev. D} {\bfseries 107} (2023) 024045} [\href{https://arxiv.org/abs/2105.04689}{{\ttfamily 2105.04689}}].


\bibitem{Sadeghi:2020ciy}
J.~Sadeghi, S.~N.~Gashti, I.~Sakalli and B.~Pourhassan, \emph{{Weak gravity conjecture of charged-rotating-AdS black hole surrounded by quintessence and string cloud}}, \href{https://doi.org/10.1016/j.nuclphysb.2024.116581}{\emph{Nucl. Phys. B} {\bfseries 1004} (2024) 116581} [\href{https://arxiv.org/abs/2011.05109}{{\ttfamily 2011.05109}}].

\bibitem{Goon:2019faz}
G.~Goon and R.~Penco, \emph{{Universal Relation between Corrections to Entropy
  and Extremality}},
  \href{https://doi.org/10.1103/PhysRevLett.124.101103}{\emph{Phys. Rev. Lett.}
  {\bfseries 124} (2020) 101103}
  [\href{https://arxiv.org/abs/1909.05254}{{\ttfamily 1909.05254}}].


\bibitem{Marrani:2022jpt}
A.~Marrani, A.~Mishra and P.~K.~Tripathy,
\emph{Non-BPS black branes in M-theory over Calabi-Yau threefolds. (Non-)uniqueness and recombination of non-BPS black strings in single modulus CICY and THCY models},
JHEP \textbf{06} (2022), 163,
doi:10.1007/JHEP06(2022)163,
[\href{https://arxiv.org/abs/2202.06872}{{\ttfamily arXiv:2202.06872 [hep-th]}}].

\bibitem{Mishra:2023ylz}
A.~Mishra and P.~K.~Tripathy,
\emph{Stable non-BPS black holes and black strings in five dimensions},
Phys. Rev. D \textbf{109} (2024) no.2, 026001,
doi:10.1103/PhysRevD.109.026001,
[\href{https://arxiv.org/abs/2310.14187}{{\ttfamily arXiv:2310.14187 [hep-th]}}].

\bibitem{Alipour:2024jgn}
M.~R.~Alipour, M.~A.~S.~Afshar, S.~Noori Gashti and J.~Sadeghi,
\emph{Weak Gravity Conjecture Validation with Photon Spheres of Quantum Corrected AdS-Reissner-Nordstrom Black Holes in Kiselev Spacetime},
[\href{https://arxiv.org/abs/2410.14352}{{\ttfamily arXiv:2410.14352}}].

\end{thebibliography}
\end{document}